\documentclass[11pt,twoside]{article}

\usepackage{asp2006}
\usepackage{epsf}
\usepackage{lscape}
\usepackage{graphicx}

\begin{document}
\title{Bent-Double Radio Sources as Probes of the Intragroup Medium}   
\author{Emily Freeland and Eric Wilcots}   
\affil{Department of Astronomy, University of Wisconsin, 475 N. Charter St., Madison, WI 53706, USA}    

\begin{abstract} 
Galaxy groups likely contain a significant fraction of the total baryonic mass in the local universe within their intragroup medium (IGM).  However, aside from a handful of UV absorption line and X-ray observations, almost nothing is known about the IGM.  We present early results from a research program that combines low-frequency radio continuum observations and optical spectroscopy of bent-double radio sources and their neighbors in groups of galaxies.  These observations allow us to probe the density of the IGM to an unprecedented degree by examining its impact on the jets of radio galaxies.
\end{abstract}



\section{Introduction}
The vast majority of galaxies, including our own Milky Way, reside in groups: small dynamical systems typically containing a handful of large ($\sim\hspace{-1ex} L_*$) galaxies and numerous smaller ones \citep{1987ApJ...321..280T}.  Groups exhibit a wide range of characteristics but tend to have dynamical masses between $10^{12}$ and $10^{14}~\mathrm{M}_{\odot}$ and diameters of $\sim$1 Mpc.  Their velocity dispersions typically range between 30 and 500$~\mathrm{km~sec}^{-1}$.  Groups occupy a key place in the hierarchical structure of the Universe with dark matter distribution intermediate between those of individual galaxies and those of giant clusters \citep{2001ApJ...563..736C}.  The baryons located in the intragroup medium (IGM) account for a significant, although highly uncertain, fraction of the local baryon density \citep{1998ApJ...503..518F}.  Understanding the distribution and evolution of this gas is an important step in studying the baryon content of the local universe. 
   
In the local universe, attempts to measure the baryon content in the form of stars, hot X-ray emitting gas, and cold gas account for less than one third of the baryon density seen at high redshift \citep{1998ApJ...503..518F}.  Recent hydrodynamical simulations predict that these missing baryons may exist in a warm-hot intergalactic medium (WHIM) with a temperature range of $10^5 < T < 10^7$ K \citep{2001ApJ...552..473D}.  Observational confirmation of the existance of the WHIM comes primarily from UV absorption lines in the spectra of low redshift quasars \citep{2000ApJ...534L...1T}.  However, UV absorption line observations are limited by their ability to probe only single sightlines, and there is some debate as to whether the absorption originates in the extended gaseous halos of individual galaxies, the IGM, or from large-scale filaments \citep{2003ApJS..146..125S,2000ApJ...534L...1T,2001ApJ...552..473D}.  In the cases where the absorption does appear to be associated with the IGM, the cause of the ionization, whether non-equilibrium cooling of hot gas or photoionization of low-density gas, is often unclear.  Assumptions about the extent of the absorbing system, the metallicity, and the intensity of the ionizing radiation field lead, in either case, to only rough estimates of the IGM density on the order of $n_H\sim10^{-4}-10^{-5}~\mathrm{cm}^{-3}$ \citep{2004AJ....127..199P,1998ASPC..143..261T}.  X-ray observations of groups are only sensitive to the high temperature component of the IGM.  They can determine the emission measure, $\int n_e^2 dl$, but without a detailed knowledge of the path length and ionization state the density cannot be pinned down. 


Another method of probing the IGM is to examine its impact on the jets of radio galaxies.  As a double lobed radio source travels through a dense intergalactic medium, ram pressure caused by its motion causes its jets to be swept back into a U shape.  Prior to \citet{1987AJ.....94..587B} it was believed that the combination of the dense intracluster medium and large galaxy velocities needed to form these bent-double radio sources could only be found in rich clusters.  This impression may have been falsely formed by a bias towards radio surveys of clusters and no comparable surveys of groups.  While many bent-doubles do reside in dense clusters, a surprising number are found in lower mass environments like groups of galaxies \citep{1978A&A....69..253E,1994ApJ...436...67V,1995AJ....110...46D,2001AJ....121.2915B}.  The existence of radio galaxies with bent jets in such low velocity dispersion environments suggests that the IGM may be more dense there than previously expected. 
 
\citet{1978A&A....69..253E} were among the first to recognize the tremendous importance of bent-double radio sources in groups of galaxies as probes of the IGM.  Their observations of radio sources NGC $6109$ and NGC $6137$ in groups, with optically derived velocity dispersions of 584 $\mathrm{km\ s}^{-1}$ and 291 $\mathrm{km\ s}^{-1}$ respectively, suggest that in these groups the IGM has a density of $5 \times 10^{-4}$ $\mathrm{cm}^{-3}$.  Ekers et al. concluded that if their results were indicative of groups in general, then a significant fraction of the baryon content of the local universe must reside in the IGM.

\section{Bent-Double Radio Sources}
\begin{figure}[!ht]
\begin{center}
\includegraphics[scale=0.7]{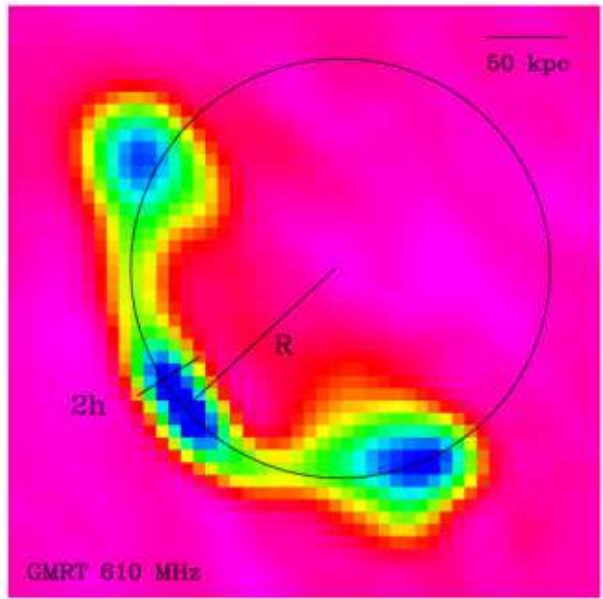}
\caption{GMRT 610 MHz continuum map (contours and grayscale) of bent-double radio source SDSS J154849.35+361035.4 labelled to illustrate the jet width, 2h, and the radius of curvature, R.  The beamsize for this dataset of $5.5^{\prime\prime}\times4.6^{\prime\prime}$ corresponds to a spatial resolution of 25 kpc $\times$ 20 kpc.}\label{fig:eqn}
\end{center}
\end{figure}

For a relativistic, hydrodynamic jet, the time-independent Euler equation describes the balance of internal and external pressure gradients:  

\begin{equation}\frac{\rho_{\mbox{\tiny IGM}} v_{gal}^2}{h}=\frac{\mathrm{w}\Gamma^2\beta^2}{R}\end{equation}  where $\rho_{\mbox{\tiny IGM}} v_{gal}^2$ is the external ram pressure, $\mathrm{w}\Gamma^2\beta^2$ is the relativistic enthalpy density inside the jet, $h$ is the radius of the jet, and $R$ is the radius of curvature of the jet.  The jet width and radius of curvature are determined observationally from the radio continuum data as illustrated in Figure \ref{fig:eqn}.  The enthalpy density is written as $\mathrm{w}=e+p$, where $e$ is the energy density and $p$ is the internal pressure.  We estimate the internal pressure in the jet using the minimum synchrotron pressure $P_{min}$ which is calculated from the radio observations of synchrotron emission at 1420 and 610 Mhz.  For the jets whose particle population is ultrarelativistic, $e=3p+\rho_{jet}c^2$, where $\rho_{jet}c^2$ is the rest mass energy density, and thus $\mathrm{w}=4P_{min} + \rho_{jet}c^2$.  While $\rho_{jet}c^2$ is not well known it is anticipated to be much less than $4P_{min}$, so we neglect this term for now.  We adopt $\beta=\frac{v}{c}=0.5$ which appears to be an average value for the jet speed in wide-angle tailed radio galaxies \citep{2006MNRAS.368..609J}.  The relativistic factor $\Gamma$ is the usual $\frac{1}{\sqrt{1-\beta^2}}$. 

\section{Sample Selection and Observations}
\begin{figure}[!ht]
\begin{center}
\includegraphics[scale=1]{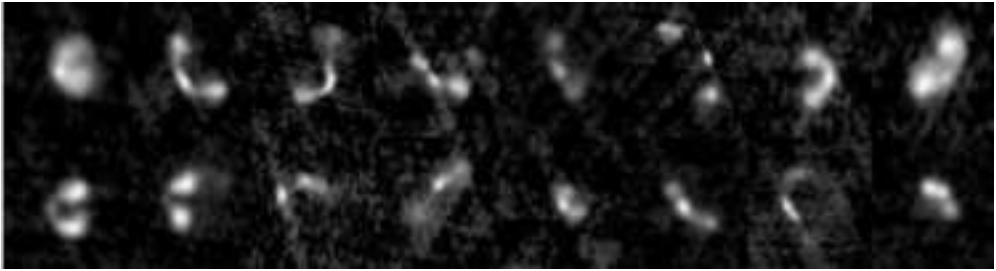}
\caption{VLA FIRST survey 1420 MHz continuum images of some of the bent-double radio sources in our final sample.}\label{fig:sample}
\end{center}
\end{figure}

We began with a sample of 839 bent-double radio galaxies identified in the VLA FIRST \footnote{Faint Images of the Radio Sky at Twenty Centimeters} survey \citep{1995ApJ...450..559B}.  Using the Sloan Digital Sky Survey (SDSS) and the NASA Extragalactic Database (NED) we were able to determine redshifts for on the order of 100 of the radio galaxies.  Sources associated with known clusters of galaxies were removed.  SDSS images for the remaining sources were examined by eye and those in regions of high galaxy density or who appeared to be interacting with neighbors were removed.  The final sample of $\sim$20 bent-double radio galaxies have redshifts in the range $0.04 < z < 0.4$.  A collection of some of the FIRST sources in our final sample are shown in Figure \ref{fig:sample}.

\citet{2001AJ....121.2915B} used broad band optical observations to estimate the richness of the environments of a sample of bent-double radio sources.  They subtracted background galaxy counts, determined from individual frames or previous deep surveys, from the total number of galaxies with absolute magnitude brighter than $M_V=-19$ within a radius of $0.5$ Mpc of the radio source.  They identify a number of radio sources that appear to be in environments with a richness less than Abell class 0.  A subset of our final sample is also included in the \citet{2001AJ....121.2915B} study and thus is already thought to exist in group environments.    

We have obtained Giant Metrewave Radio Telescope (GMRT) 610 Mhz radio continuum observations of a one degree field centered on each source.  The resolution of our GMRT observations is well matched with the FIRST 1420 MHz VLA radio continuum survey resolution.  When combined, these observations allow us to calculate the spectral index of the jets and their minimum synchrotron pressure.  We also use the radio data to measure the size of the jets and their radius of curvature.  We are currently using the WIYN Consortium\footnote{University of Wisconsin, Indiana University, Yale University, and the National Optical Astronomy Observatory} 3.5m telescope on Kitt Peak to perform multi-object spectroscopy of sources whose photometric redshifts indicate that they may belong to the same group as the radio source.  Once we have information on the group membership, we can calculate the group velocity dispersion and estimate the relative space velocity of the radio source.  Where available we use archival X-ray observations of these fields to measure or place upper limits on the temperature of the intragroup gas. 

\begin{figure}[!ht]
\begin{center}
\includegraphics*[width=2in,height=2in]{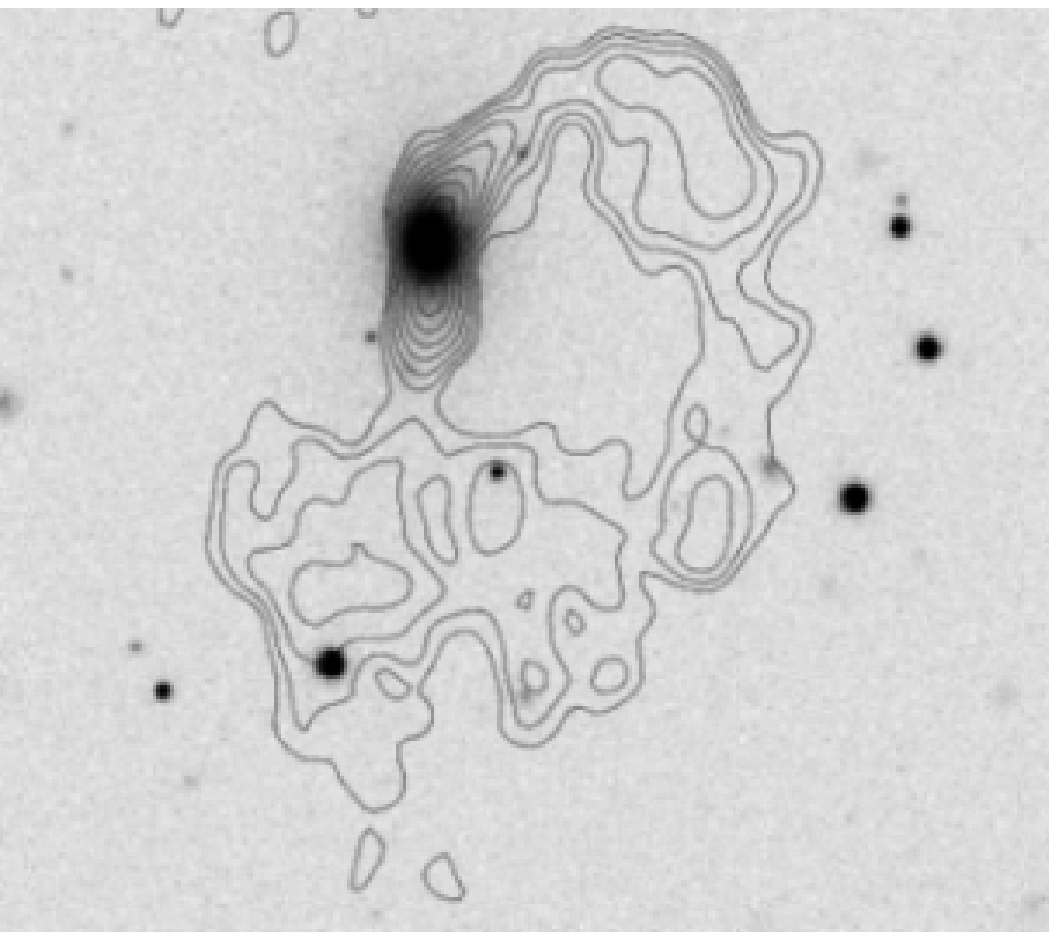}\includegraphics*[width=2in,height=2in,angle=90]{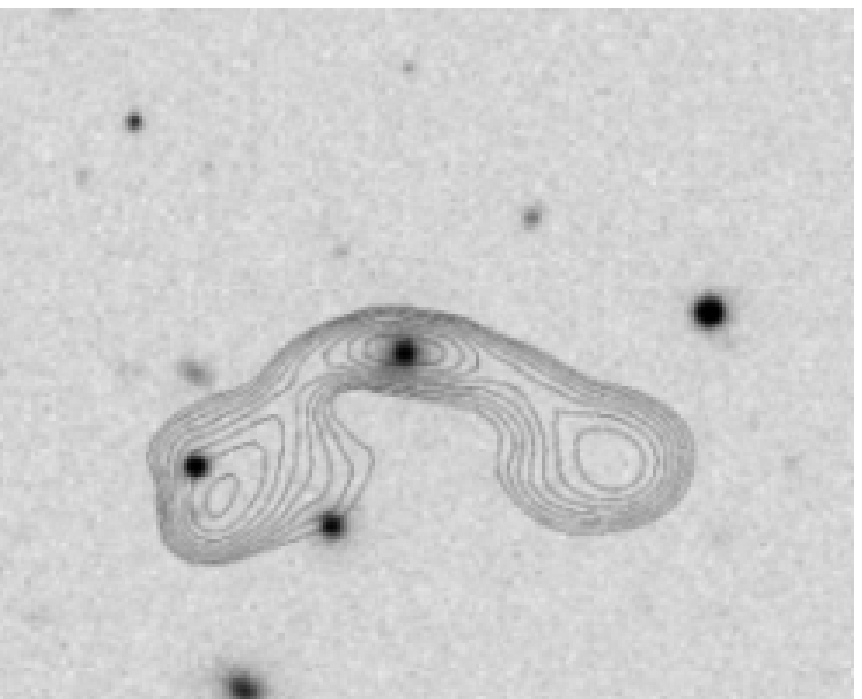}\\
\includegraphics*[width=2in,height=2in]{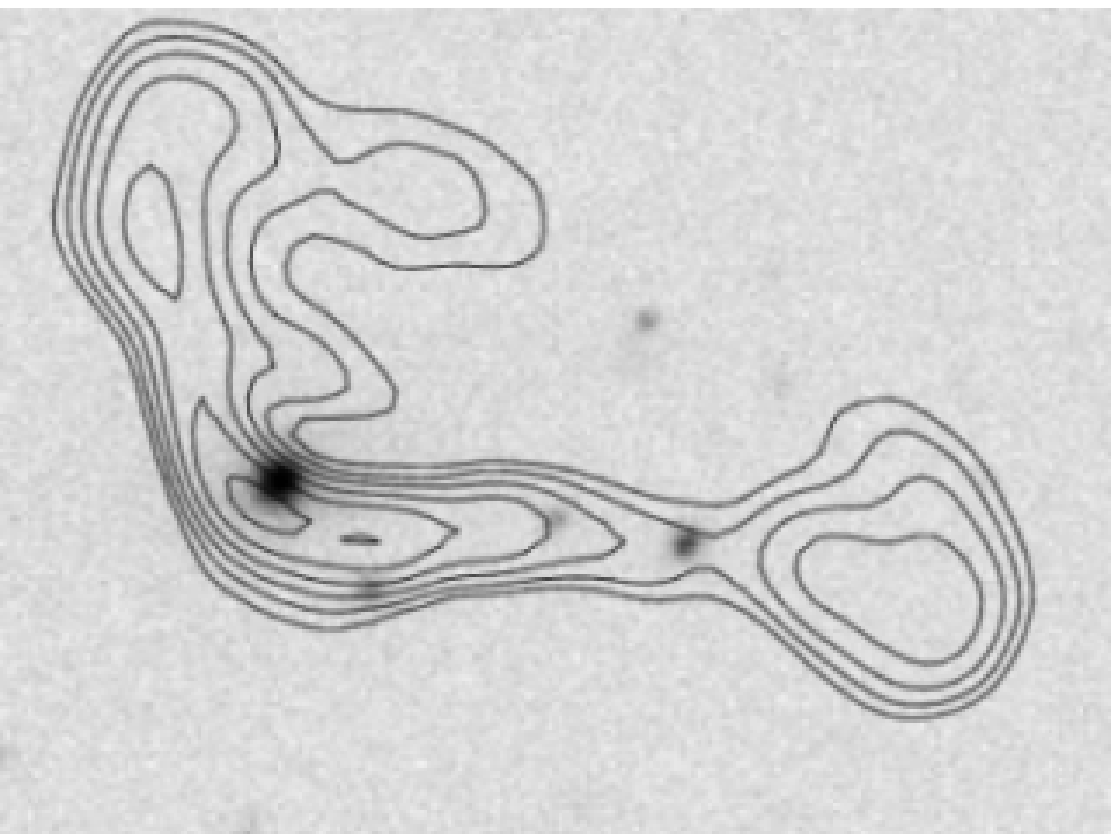}\includegraphics*[width=2in,height=2in]{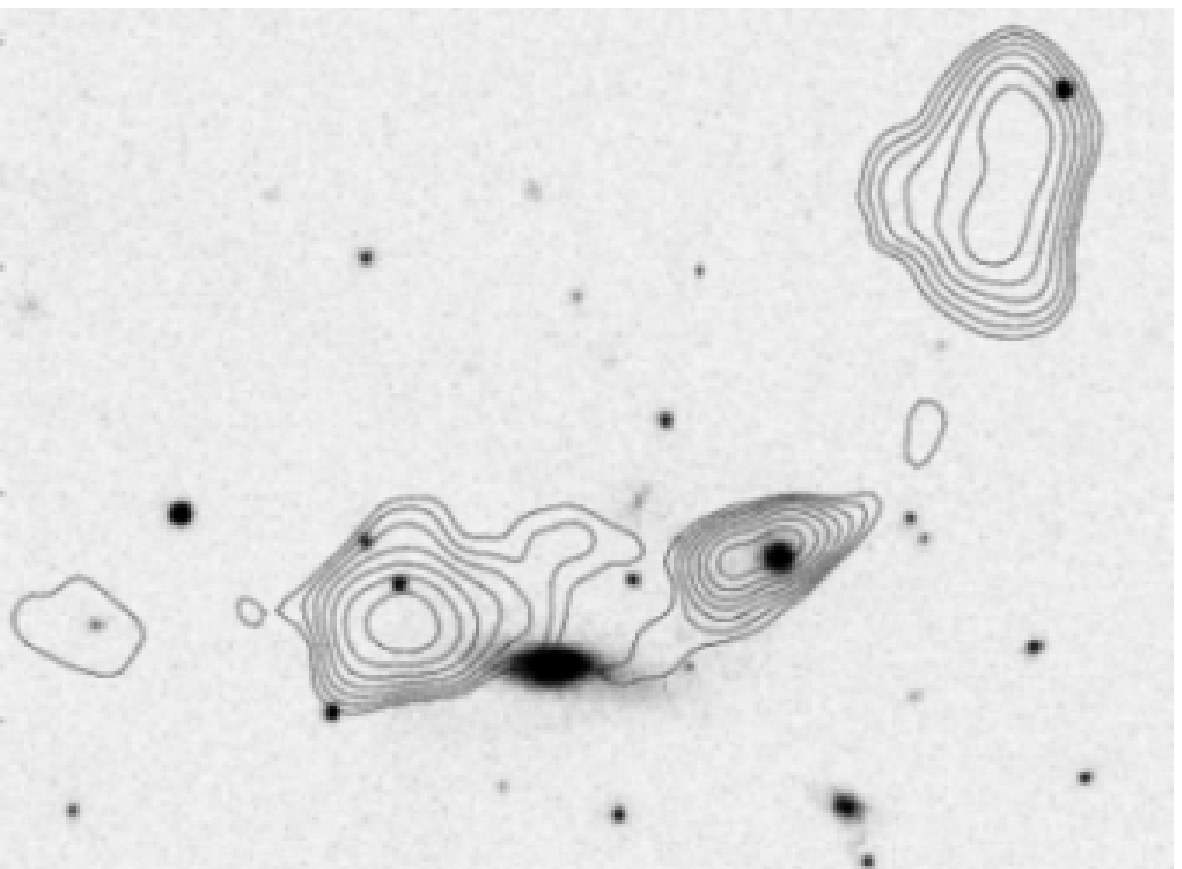}
\caption{GMRT 610 MHz contours overlaid on SDSS r band optical images for the sources listed in Table 1, shown in the same order starting clockwise from the upper left.}
\end{center}
\end{figure}


\section{Results}

We have calculated IGM densities using the four bent-double sources listed in Table 1 below.  Without the multi-object spectroscopy for the fields surrounding these sources, we do not have detailed information on individual group velocity dispersions.  \citet{1987ApJ...319..671S} give the velocity dispersion of the poor group, of which CGCG 118-054 is a member, as $300~ \mathrm{km~sec}^{-1}$ from seven member velocities.  \citet{2001AJ....121.2915B} estimate the richness of the environment around FIRST J124942.2+303838 and FIRST J112038.5+291234 as less than Abell class 0.  We assume a velocity dispersion, and corresponding radio source velocity, of $500~ \mathrm{km~sec}^{-1}$ for the remaining three sources since this is a typical value for an Abell cluster of richness class 0 \citep{1993ApJ...415L..17L}.

\renewcommand{\arraystretch}{1.25}
\begin{table}[!ht]
\caption{Source Properties and IGM Densities}
\smallskip
\begin{center}
\begin{tabular}{lcccc}
\hline
Source & z & R    & $P_{min,jet}$ & $n_{\mbox{\tiny IGM}}$ \\
       &   &(kpc) &($\mathrm{dynes~cm}^{-2}$) &($\mathrm{cm}^{-3}$)\\
\hline
CGCG 118-054             & 0.04 & 21.4 & $4\times10^{-12}$  & $9\times10^{-4}$   \\
SDSS J154849.35+361035.4 & 0.23 & 80.7 & $2\times10^{-12}$  & $2\times10^{-4}$  \\
FIRST J124942.2+303838   & 0.19 & 52.7 & $1\times10^{-12}$  & $2\times10^{-4}$  \\
FIRST J112038.5+291234   & 0.24 & 214 &  $6\times10^{-13}$  & $2\times10^{-5}$ \\
\hline
Note: $n_{\mbox{\tiny IGM}}=\frac{\rho_{\mbox{\tiny IGM}}}{\mu m_{H}}, \mu=0.6$.
\end{tabular}
\end{center}
\end{table}

The values given in Table 1 are robust lower limits on the density of the IGM in these four groups of galaxies.  If the speed of the jet is greater than $0.5c$, or the velocity of the radio galaxy is less than $500~ \mathrm{km~sec}^{-1}$, or $\rho_{jet}c^2$ contributes a significant amount of pressure, then the jet will be harder to bend and the measured density will increase.  These IGM densities are consistent with estimates from UV absorption line and X-ray observations \citep{1998AJ....115..436S,1993ApJ...404L...9M}.  Since the group environment is the most common in the local universe, even a relatively modest IGM density will contribute significantly to the local baryon density.  As we accumulate more optical data and are better able to characterize the sizes and velocity dispersions of these groups of galaxies, we will be able to quantitatively assess the contribution of the baryons in the IGM to the baryon density in the local universe.

\acknowledgements 
E.F. would like to acknowledge the generous support of the Wisconsin Space Grant Consortium.  This research has made use of the NASA/IPAC Extragalactic Database (NED) which is operated by the Jet Propulsion Laboratory, California Institute of Technology, under contract with the National Aeronautics and Space Administration.  We thank the staff of the GMRT who have made these observations possible. GMRT is run by the National Centre for Radio Astrophysics of the Tata Institute of Fundamental Research.


\bibliographystyle{apj}
\bibliography{freeland}

\end{document}